\documentclass[aps,prd,superscriptaddress,nofootinbib,amsmath,amsfonts,preprintnumbers,groupedaddress,showpacs,9pt,english]{revtex4-1}
\usepackage{amsmath}
\usepackage{amssymb}
\usepackage{babel}
\usepackage{wrapfig}
\usepackage{cancel}
\usepackage{float}

\usepackage{graphicx}
\graphicspath{{figures/}}
\usepackage{amsmath,amssymb}  
\usepackage{mathrsfs}         
\usepackage[colorlinks,citecolor=blue,urlcolor=blue,linkcolor=blue]{hyperref}
\usepackage[figtopcap]{subfigure}
\usepackage{color}
\usepackage{relsize,exscale}
\makeatletter
\newcommand{\beq}{\begin{equation}}
\newcommand{\eeq}{\end{equation}}
\newcommand{\bea}{\begin{eqnarray}}
\newcommand{\eea}{\end{eqnarray}}

\newcommand{\comm}[1]{}

\usepackage{array,multirow,graphicx}
\usepackage{dcolumn}
\usepackage{newlfont}
\usepackage{bm}
\usepackage[colorlinks,citecolor=blue,urlcolor=blue,linkcolor=blue]{hyperref}
\usepackage[figtopcap]{subfigure}
\usepackage{color}
\usepackage{amsmath,amssymb,amsthm,graphicx,hyperref,geometry}
\usepackage{xcolor}
\usepackage{booktabs}

\usepackage{scalerel}
\usepackage{tikz}
\usetikzlibrary{svg.path}
\definecolor{orcidlogocol}{HTML}{A6CE39}
\tikzset{
  orcidlogo/.pic={
    \fill[orcidlogocol] svg{M256,128c0,70.7-57.3,128-128,128C57.3,256,0,198.7,0,128C0,57.3,57.3,0,128,0C198.7,0,256,57.3,256,128z};
    \fill[white] svg{M86.3,186.2H70.9V79.1h15.4v48.4V186.2z}
                 svg{M108.9,79.1h41.6c39.6,0,57,28.3,57,53.6c0,27.5-21.5,53.6-56.8,53.6h-41.8V79.1z M124.3,172.4h24.5c34.9,0,42.9-26.5,42.9-39.7c0-21.5-13.7-39.7-43.7-39.7h-23.7V172.4z}
                 svg{M88.7,56.8c0,5.5-4.5,10.1-10.1,10.1c-5.6,0-10.1-4.6-10.1-10.1c0-5.6,4.5-10.1,10.1-10.1C84.2,46.7,88.7,51.3,88.7,56.8z};}}
\newcommand\orcid[1]{\href{https://orcid.org/#1}{\mbox{\scalerel*{
\begin{tikzpicture}[yscale=-1,transform shape]
\pic{orcidlogo};
\end{tikzpicture}
}{|}}}}
\begin{document}
\title{\textbf{Primordial Black Hole Formation in
$f(R)=R+\alpha R^2$ Gravity: \\
Perturbative and Non--Perturbative Analysis}}

\author{G.~G.~L.~Nashed$^{1}$~\orcid{0000-0001-5544-1119}}
\email{nashed@bue.edu.eg}
\author{A.~Eid$^{2}$}\email{amaid@imamu.edu.sa}
\affiliation {$^{1}$ Centre for Theoretical Physics, The British University, P.O. Box
43, El Sherouk City, Cairo 11837, Egypt\\$^{2}$Department of Physics, College of Science, Imam Mohammad Ibn Saud Islamic University (IMSIU), Riyadh, Kingdom of Saudi Arabia, }


\begin{abstract}
We present a complete analytic and semi-analytic study of gravitational collapse and primordial black hole (PBH) formation in the quadratic
$f(R)$ model $f(R)=R+\alpha R^2$. We first derive the perturbative expansion around General Relativity (GR),
working to first order in the small parameter $\alpha$. For a collapsing flat FLRW dust interior we compute the explicit
first-order corrections to the scale factor, the stellar radius, and the horizon formation time.
{ We then use these results to infer the expected trend in the PBH formation threshold
$\delta_c$, rather than a direct quantitative determination.
  Within this perturbative framework, the quadratic correction modifies the dust collapse dynamics at first order, while the flat radiation-dominated background does not receive a nontrivial correction at the same order. As result, any modification of PBH formation in the radiation era must arise from nonlinear or non-perturbative effects.
 The perturbative analysis therefore provides qualitative insight into how curvature corrections influence collapse, particularly in high-curvature regimes.}
To access this regime we reformulate the theory
in the Einstein frame, where the model becomes GR plus the scalaron field
$\phi$ with the Starobinsky potential.
We provide the complete ODE system governing both the cosmological
background and the evolution of an overdense closed FLRW patch.
This system can be numerically integrated to obtain the critical
overdensity $\delta_c(k)$ for PBH formation near the end of inflation.
\end{abstract}
\maketitle

\tableofcontents

\section{Introduction}

Black holes are among the most fundamental and robust predictions of
General Relativity (GR).
The earliest exact vacuum solution with spherical symmetry is the Schwarzschild
solution \cite{Schwarzschild:1916uq}, extended to charged black holes by
Reissner and Nordstr\"om \cite{Reissner:1916cle,Nordstrom:2018acn} and to rotating
solutions by Kerr \cite{Kerr:1963ud}.  These geometries are characterized
by event horizons, curvature singularities, and well-defined causal structure
\cite{Hawking:1973uf,Wald:1984rg}.

The uniqueness theorems (often summarized as ``no-hair'' theorems) state that
stationary, asymptotically flat vacuum black holes in GR are described
completely by mass, charge, and angular momentum \cite{Robinson:1975bv,Nashed:2021gkp,Heusler:1996jaf}.
The laws of black hole mechanics, discovered by
Bardeen, Carter, and Hawking \cite{Bardeen:1973gs}, together with Hawking's
semiclassical radiation result \cite{Hawking:1974rv,Hawking:1975vcx},
establish black holes as thermodynamic objects with temperature and entropy.

Observations of black holes span a wide range of masses, from stellar-mass
systems in X-ray binaries and gravitational-wave events
\cite{LIGOScientific:2016emj,KAGRA:2021vkt} to supermassive black holes in galactic
centers \cite{EventHorizonTelescope:2019dse,EventHorizonTelescope:2022wkp}.
These observations provide strong tests of strong-field GR
and constraints on alternative theories of gravity.

The class of $f(R)$ theories generalizes the Einstein--Hilbert action of General Relativity (GR) by replacing the Ricci scalar $R$ with a nonlinear function $f(R)$. This modification is motivated by the expectation that classical GR may receive corrections in regimes of high or low curvature, for instance in the early universe, near compact objects, or on cosmological scales. In many cases these theories can mimic dark energy or drive inflation purely from geometry, without introducing additional fundamental matter fields. For comprehensive reviews, see e.g.\ De Felice and Tsujikawa \cite{DeFelice:2010aj}, Sporea \cite{Sporea:2014apa}, and more recent cosmological and gravitational-wave oriented discussions in Refs.~\cite{Bajardi:2022ocw,Dejrah:2025byj}.

In modified gravity theories, including $f(R)$ gravity, black hole solutions
may differ from GR or in some cases coincide with the standard solutions.
For many $f(R)$ models, vacuum solutions with $R=0$ remain solutions, so the
Schwarzschild solution often persists \cite{DeFelice:2010aj}.
However, more general solutions with nonzero curvature,
scalar ``hair'', or modified horizon structures may arise
when the scalaron becomes dynamical or couples to matter in nontrivial ways
\cite{Sotiriou:2008rp,Nashed:2023pxd,Capozziello:2011et}.
Stability analyses, scalar perturbations, and matching of interior and exterior
solutions remain active research topics.

Black hole thermodynamics and Hawking radiation may also be modified in
$f(R)$ gravity through changes in entropy relations, effective
Newton constant, or quantum corrections \cite{Cognola:2005de,Nashed:2023pxd,Briscese:2006xu}.
In strong curvature regimes (e.g.\ near singularities), higher-order curvature
terms may regularize or alter the classical structure of black holes, though
viable, stable, and observationally consistent solutions are difficult to obtain.

Primordial black holes (PBHs) are black holes formed in the early universe
from the gravitational collapse of overdense regions \cite{Hawking:1971ei,Carr:1974nx}.
Unlike astrophysical black holes, PBHs may form with masses spanning
a vast range $
10^{-5}\,\mathrm{g} \ \lesssim M_{\textrm PBH} \lesssim 10^{5} M_{\odot}$
depending on the epoch of formation \cite{Carr:2020gox}.

The dominant formation mechanism is the collapse of large density fluctuations
when they re-enter the Hubble horizon during the radiation-dominated era
\cite{Carr:1975qj}.
If the density contrast at horizon re-entry exceeds a critical threshold
$\delta > \delta_c$, the region collapses into a black hole.
The threshold depends on the equation of state and the shape of the perturbation
\cite{Musco:2018rwt,ElHanafy:2014efn,Harada:2013epa}.
For radiation domination, numerical studies find
$\delta_c \approx 0.4$--$0.5$. Alternative PBH formation mechanisms include:
\begin{itemize}
    \item Collapse of cosmic strings or domain walls \cite{Hawking:1987bn};
    \item Bubble collisions and first-order phase transitions \cite{Hawking:1982ga};
    \item Scalar field fragmentation or preheating \cite{Jedamzik:1998hc,Green:2020jor};
    \item Models with modified gravity or enhanced curvature fluctuations
    \cite{Motohashi:2017kbs,Ozsoy:2023ryl}.
\end{itemize}

In $f(R)=R+\alpha R^2$ gravity (the Starobinsky model), enhanced scalaron
dynamics during the end of inflation or reheating can modify the collapse
threshold, potentially making PBH formation easier under certain conditions
\cite{Motohashi:2017kbs}.
The scalar degree of freedom changes the effective equation of state and
the growth of curvature perturbations, influencing PBH abundance.

In the perturbative regime ($\alpha R\ll1$), PBH formation resembles GR closely,
with only small corrections to collapse times or thresholds.
However, in the high-curvature regime near the end of inflation, full
non-perturbative scalar-tensor dynamics can significantly alter PBH production
\cite{Ozsoy:2023ryl}.

Black holes are among the most fundamental predictions of General Relativity (GR),
arising as exact solutions such as the Schwarzschild, Reissner--Nordstr\"om, and Kerr geometries.
They play a central role in modern gravitational physics, astrophysics, and cosmology.
However, in regimes of extremely high curvature, near singularities, during gravitational
collapse, or in the early universe, quantum or semiclassical corrections to GR are expected
to become significant. This provides strong motivation for exploring modified theories of
gravity that incorporate higher-order curvature terms.

Among such theories, the quadratic model of $f(R)$
known as the Starobinsky model, is of particular interest. It introduces a single scalar
degree of freedom (the scalaron), remains ghost-free, and famously provides a successful
inflationary scenario. Therefore, this model naturally connects early-universe cosmology
with strong-gravity physics.

Primordial black holes (PBHs) are black holes formed in the early universe from the collapse
of overdense regions. Their formation depends sensitively on the dynamics of gravitational
collapse and on the critical overdensity threshold $\delta_c$ at horizon re-entry. In GR,
$\delta_c$ is well studied, but in modified gravity theories the collapse dynamics may be
altered, potentially shifting the horizon formation time and the value of $\delta_c$.

The central motivation of this study is to analyze gravitational collapse and PBH formation
in the context of the quadratic $f(R)$ model. Specifically, we aim to:
\begin{itemize}
    \item determine how curvature corrections modify the collapse of a dust FLRW region,
    \item compute the shift in horizon formation time relative to GR,
    \item { assess how the modified collapse dynamics may affect the PBH formation threshold $\delta_c$,}
    \item and formulate the full non-perturbative dynamics in the Einstein frame, suitable
          for numerical PBH formation studies near the end of Starobinsky inflation.
\end{itemize}
These results offer insight into whether modified gravity can enhance or suppress PBH
production and how strong-curvature corrections affect the formation of compact objects.
\vspace{0.3cm}

The structure of the study as follows: In Sec.~\ref{fR}  we introduce the action, field equations, scalar--tensor equivalence, and relevant
mathematical tools of $f(R)$ gravity.  In Sec.~\ref{prt} perturbative expansion around GR of the field equations has been discussed. In Sec.~\ref{frw} we calculate how the collapse time and horizon formation differ from GR and shows that $\alpha>0$ accelerates collapse.   In Sec.~\ref{ef} we describes the full scalaron dynamics and provides the complete ODE system for both the background and a closed, overdense FLRW patch. { Section~\ref{frw1} relates the modified collapse time to the expected trend of the PBH threshold within the perturbative framework, while Sec.~\ref{ef} provides the nonlinear Einstein-frame formulation required for a quantitative determination of $\delta_c^{(\alpha)}(k)$.}
\section{Basic Formulation of $f(R)$ Gravity}\label{fR}

\subsection{Action and Field Equations}

In the \emph{metric} formalism, the action of $f(R)$ gravity is
\begin{equation}\label{act}
S = \frac{1}{2\kappa^2}\int d^4x\,\sqrt{-g}\,f(R) + S_m[g_{\mu\nu},\Psi],
\end{equation}
where $\kappa^2 = 8\pi G$, $g$ is the determinant of the metric, and $S_m$ is the matter action depending on generic matter fields $\Psi$.

Varying the action \eqref{act} with respect to the metric yields the modified Einstein equations
\begin{equation}\label{fr2}
f_R R_{\mu\nu} - \frac{1}{2}f g_{\mu\nu}
+ \left(g_{\mu\nu}\Box - \nabla_\mu\nabla_\nu\right)f_R
= \kappa^2 T_{\mu\nu},
\end{equation}
where $f_R \equiv df/dR$, $\Box = g^{\rho\sigma}\nabla_\rho\nabla_\sigma$ is the d'Alembertian, and $T_{\mu\nu}$ is the energy--momentum tensor. Equations \eqref{fr2} can be recast as
\begin{equation}
G_{\mu\nu} + K_{\mu\nu} = \kappa^2 T_{\mu\nu},
\end{equation}
where $G_{\mu\nu}$ is the usual Einstein tensor and $K_{\mu\nu}$ encodes the corrections from the nonlinear function $f(R)$.
In general, $f(R)$ theories propagate an extra scalar degree of freedom (the ``scalaron'') in addition to the two tensor modes of GR. It is well-known that metric $f(R)$ theories are dynamically equivalent to a subclass of scalar--tensor theories where the action \eqref{act} can be rewritten as
\begin{equation}
S = \frac{1}{2\kappa^2}\int d^4x\sqrt{-g}\,\big[f'(\chi)(R - \chi) + f(\chi)\big] + S_m,
\end{equation}
whose variation with respect to $\chi$ gives $\chi = R$.
Defining $f'(\chi)\equiv \Phi$ and performing a conformal transformation to the Einstein frame,
$g^E_{\mu\nu} = \Phi\,g_{\mu\nu}$,
one obtains GR plus a canonical scalar field $\phi$ with a potential $V(\phi)$ determined by the original $f(R)$ function \cite{Sporea:2014apa,DeFelice:2010aj}. This equivalence is extremely useful for studying stability, cosmological dynamics, and perturbations in $f(R)$ gravity. In addition to the metric formalism, one can formulate $f(R)$ theories in the Palatini approach, where the connection is treated as an independent variable.
The resulting dynamics differ significantly from the metric case; in Palatini $f(R)$ the scalar degree of freedom becomes non-dynamical, and the field equations are second order.
There are also more general metric--affine approaches where both metric and connection are varied independently and allowed to couple in more complicated way, for more details see ~\cite{DeFelice:2010aj,Nashed:2021pah}.

\section{Perturbative Expansion Around GR}\label{prt}

In this section we derive the linearized field equations for the
$f(R)$ model around a GR background.  We work in the metric
formalism.
\subsection{Field Equations of the Quadratic Model and Expansion Around a GR Background}
Substituting the quadratic form of $f(R) = R + \alpha R^2$  into \eqref{fr2}  we get:
\begin{align}\label{fr3}
 (1+2\alpha R) R_{\mu\nu}
 &- \frac12 (R+\alpha R^2) g_{\mu\nu}
 + \left(g_{\mu\nu}\Box - \nabla_\mu\nabla_\nu\right)(1+2\alpha R)
 = \kappa^2 T_{\mu\nu}.
\end{align}
We now expand around a GR background solution $g^{(0)}_{\mu\nu}$ that
satisfies the Einstein equations
\begin{equation}
G^{(0)}_{\mu\nu} \equiv G_{\mu\nu}\big[g^{(0)}\big] = \kappa^2 T_{\mu\nu}.
\label{eq:GR-background}
\end{equation}
The metric, Ricci tensor, and Ricci scalar are expanded as
\begin{equation}
g_{\mu\nu} = g^{(0)}_{\mu\nu} + \alpha h_{\mu\nu},\qquad
R_{\mu\nu} = R^{(0)}_{\mu\nu} + \alpha R^{(1)}_{\mu\nu} + \mathcal{O}(\alpha^2),
\qquad
R = R^{(0)} + \alpha R^{(1)} + \mathcal{O}(\alpha^2),
\end{equation}
and similarly for the Einstein tensor,
\begin{equation}
G_{\mu\nu} = G^{(0)}_{\mu\nu} + \alpha G^{(1)}_{\mu\nu}[h]
+ \mathcal{O}(\alpha^2).
\end{equation}
Here $G^{(1)}_{\mu\nu}[h]$ denotes the linear response of $G_{\mu\nu}$ to
the perturbation $h_{\mu\nu}$, evaluated on the background $g^{(0)}_{\mu\nu}$.

Because the $K_{\mu\nu}$ term in \eqref{fr3} is already multiplied
by an explicit factor of $\alpha$, its contribution at order $\alpha$
comes purely from evaluating $K_{\mu\nu}$ on the background metric:
\begin{equation}
K_{\mu\nu} = K^{(0)}_{\mu\nu} + \mathcal{O}(\alpha),
\qquad
K^{(0)}_{\mu\nu} \equiv
2 R^{(0)} R^{(0)}_{\mu\nu}
- \frac{1}{2} (R^{(0)})^2 g^{(0)}_{\mu\nu}
+ 2 \bigl(g^{(0)}_{\mu\nu}\Box^{(0)} R^{(0)} - \nabla^{(0)}_\mu\nabla^{(0)}_\nu R^{(0)}\bigr).
\end{equation}
Any corrections to $K_{\mu\nu}$ that are linear in $h_{\mu\nu}$ would appear
multiplied by $\alpha$ from the expansion of $K_{\mu\nu}$ \emph{and}
another factor of $\alpha$ from the prefactor in \eqref{fr3}, and
hence would be $\mathcal{O}(\alpha^2)$; we neglect them in a first-order
treatment.

\subsection{Linearized Field Equations}

Substituting the expansions into \eqref{fr3} we obtain
\begin{align}
G_{\mu\nu} + \alpha K_{\mu\nu}
&= G^{(0)}_{\mu\nu} + \alpha\Bigl( G^{(1)}_{\mu\nu}[h] + K^{(0)}_{\mu\nu}\Bigr)
 + \mathcal{O}(\alpha^2).
\end{align}
The matter energy--momentum tensor $T_{\mu\nu}$ is the same as in the
background if we assume that matter is minimally coupled and we do not
add an explicit perturbation to $T_{\mu\nu}$ at this stage, so
\begin{equation}
\kappa^2 T_{\mu\nu} = \kappa^2 T_{\mu\nu}^{(0)}.
\end{equation}
Using the background Einstein equation \eqref{eq:GR-background},
\begin{equation}
G^{(0)}_{\mu\nu} = \kappa^2 T_{\mu\nu}^{(0)},
\end{equation}
we find that the $\mathcal{O}(1)$ part of the field equations is automatically
satisfied.  At order $\alpha$ we obtain the linearized equation
\begin{equation}
G^{(1)}_{\mu\nu}[h] + K^{(0)}_{\mu\nu} = 0.
\end{equation}
It is convenient to denote
\begin{equation}
\delta^{(0)}_{\mu\nu}[g^{(0)}] \equiv K^{(0)}_{\mu\nu},
\end{equation}
which is a functional of the background metric only.  The master perturbation
equation can then be written as
\begin{equation}
G^{(1)}_{\mu\nu}[h] + \delta^{(0)}_{\mu\nu}[g^{(0)}] = 0.
\label{eq:pert-master}
\end{equation}
This is the desired equation: the linearized Einstein tensor of the metric
perturbation $h_{\mu\nu}$ is sourced by the curvature corrections evaluated
on the GR background.


For completeness we recall the standard expressions for the first-order
variations of the curvature tensors.  The perturbation of the inverse metric
is
\begin{equation}
\delta g^{\mu\nu} = - h^{\mu\nu},
\qquad
h \equiv g^{(0)\mu\nu} h_{\mu\nu},
\end{equation}
and the variation of the Christoffel symbols is
\begin{equation}
\delta \Gamma^\rho_{\mu\nu}
= \frac12 g^{(0)\rho\sigma}
\Bigl(\nabla^{(0)}_\mu h_{\sigma\nu}
     + \nabla^{(0)}_\nu h_{\sigma\mu}
     - \nabla^{(0)}_\sigma h_{\mu\nu}\Bigr).
\end{equation}
From this one obtains the variation of the Ricci tensor,
\begin{align}
R^{(1)}_{\mu\nu}
= \delta R_{\mu\nu}
&= \nabla^{(0)}_\rho \delta\Gamma^\rho_{\mu\nu}
 - \nabla^{(0)}_\nu \delta\Gamma^\rho_{\mu\rho} \\[2mm]
&= \frac12\Bigl(
 - \Box^{(0)} h_{\mu\nu}
 - \nabla^{(0)}_\mu\nabla^{(0)}_\nu h
 + \nabla^{(0)}_\rho\nabla^{(0)}_\mu h^\rho{}_{\nu}
 + \nabla^{(0)}_\rho\nabla^{(0)}_\nu h^\rho{}_{\mu}
\Bigr)
 + R^{(0)}_{\rho(\mu}h^\rho{}_{\nu)}
 - R^{(0)}_{\rho\mu\sigma\nu} h^{\rho\sigma}, \nonumber
\end{align}
and the variation of the Ricci scalar,
\begin{equation}
R^{(1)} = \delta R
= - R^{(0)}_{\mu\nu} h^{\mu\nu}
+ \nabla^{(0)}_\mu \nabla^{(0)}_\nu h^{\mu\nu}
- \Box^{(0)} h.
\end{equation}
The linearized Einstein tensor
\begin{equation}
G^{(1)}_{\mu\nu}[h] = R^{(1)}_{\mu\nu}
- \frac12 g^{(0)}_{\mu\nu} R^{(1)}
- \frac12 h_{\mu\nu} R^{(0)}
+ \frac12 g^{(0)}_{\mu\nu} h^{\rho\sigma} R^{(0)}_{\rho\sigma}
\end{equation}
is obtained by combining these expressions.
In practical applications (such as the FLRW interior considered later) the
background symmetries simplify $G^{(1)}_{\mu\nu}[h]$ considerably, but the
master equation \eqref{eq:pert-master} is valid for any GR background
$g^{(0)}_{\mu\nu}$.

{
We emphasize that this homogeneous treatment is not intended to provide a direct determination of the PBH formation threshold $\delta_c$. Rather, it serves as a simplified framework to isolate how curvature corrections modify the background collapse dynamics. The connection to $\delta_c$ will therefore be discussed only at a qualitative level.
}

\section{Collapsing Flat FLRW Interior}\label{frw}
{
In this section we employ a homogeneous collapse approximation as a simplified proxy to explore the effect of curvature corrections on gravitational collapse. This approach neglects spatial inhomogeneities and should therefore be interpreted as providing qualitative insight rather than a first-principles description of PBH formation. A quantitative treatment requires the nonlinear evolution of an inhomogeneous overdense region, as discussed in Sec.~\ref{ef}. We emphasize that this homogeneous treatment is not intended to provide a direct determination of the PBH formation threshold $\delta_c$, but rather to isolate how curvature corrections modify the background collapse dynamics. Consequently, any connection to $\delta_c$ should be understood as qualitative and heuristic. This approximation should be understood as a perturbative, homogeneous description and not as a realistic model of PBH formation.
}

To investigate the effect of the $R^2$ correction on gravitational collapse, we
consider a spatially flat FLRW interior metric of the form
\begin{equation}
ds^2 = -dt^2 + a^2(t)\left(dr^2+r^2 d\Omega^2\right),
\label{eq:FLRWmetric}
\end{equation}
filled with comoving dust ($p=0$).
This metric describes the interior region of a spherically symmetric collapsing
cloud in the Oppenheimer--Snyder picture. In General Relativity, the dynamics of a dust FLRW universe are governed by the
standard Friedmann equation,
\begin{equation}
3H_0^2 = \kappa^2 \rho,
\qquad H_0 \equiv \frac{\dot a_0}{a_0}.
\end{equation}
Since dust satisfies $\rho \propto a_0^{-3}$, the solution for a collapsing branch
can be written as
\begin{equation}
a_0(t) = A (t_0 - t)^{2/3},
\qquad u \equiv t_0 - t,
\label{eq:a0solution}
\end{equation}
where $t_0$ denotes the time at which the scale factor collapses to zero and $A$ is a constant fixed by initial conditions.

{The flat FLRW metric employed here should be interpreted as describing the homogeneous interior region of a collapsing overdensity, in analogy with the Oppenheimer--Snyder construction. It does not by itself constitute a complete Oppenheimer--Snyder collapse model, which would require matching to an exterior vacuum solution. Accordingly, the present treatment should be viewed as a simplified description capturing the homogeneous dynamics.}

From Eq.~\eqref{eq:a0solution}, the Hubble parameter becomes
\begin{equation}
H_0 = \frac{\dot a_0}{a_0}
    = \frac{-\frac{2}{3}A u^{-1/3}}{A u^{2/3}}
    = -\frac{2}{3}u^{-1}.
\label{eq:H0}
\end{equation}
The negative sign reflects the fact that the background is collapsing.


The Ricci scalar for a flat FLRW metric is given by
\[
R = 6(\dot H + 2H^2).
\]
Using $H_0$ from \eqref{eq:H0}, we compute
\begin{equation}
\dot H_0 = -\frac{2}{3}(-1)u^{-2} = -\frac{2}{3} u^{-2},
\end{equation}
and therefore  the GR background curvature yields:
\begin{equation}
R^{(0)} = 6\left(-\frac{2}{3}u^{-2} + 2\frac{4}{9}u^{-2}\right)
        = \frac{4}{3} u^{-2}.
\label{eq:R0}
\end{equation}
Differentiating yields
\begin{equation}
\dot R^{(0)} = \frac{d}{dt}\!\left(\frac{4}{3}u^{-2}\right)
             = \frac{8}{3}u^{-3}.
\label{eq:R0dot}
\end{equation}


In $f(R)=R+\alpha R^2$ gravity, the correction term in the modified Einstein
equations is
\[
K_{\mu\nu} =
2R R_{\mu\nu}
- \frac12 R^2 g_{\mu\nu}
+ 2\left(g_{\mu\nu}\Box R - \nabla_\mu\nabla_\nu R\right).
\]

Since the background is homogeneous, all spatial derivatives of $R^{(0)}$ vanish,
and only time derivatives contribute.
A direct evaluation using \eqref{eq:R0}--\eqref{eq:R0dot} yields
\begin{equation}
K^{(0)}_{00} = - 8 u^{-4}.
\label{eq:K00}
\end{equation}
This term acts as a ``source'' for the first-order perturbation induced by the
$R^2$ modification. Now we expand the Hubble parameter as
\[
H = H_0 + \alpha H_1 + \mathcal{O}(\alpha^2),
\]
and linearize the modified Friedmann equation.
Using the GR result $3H_0^2 = \kappa^2 \rho$ and keeping only the terms of order
$\alpha$ gives the first-order constraint:
\begin{equation}
6H_0 H_1 + K^{(0)}_{00} = 0.
\label{eq:linearFriedmann}
\end{equation}

Substituting $H_0$ from \eqref{eq:H0} and $K^{(0)}_{00}$ from \eqref{eq:K00}, we find:
\begin{equation}
6\left(-\frac{2}{3}u^{-1}\right)H_1 - 8u^{-4} = 0, \quad \mbox{which  yields} \quad
H_1 = -2 u^{-3}.
\label{eq:H1}
\end{equation}
We write the perturbed scale factor in the form
\begin{align}
a(t) = a_0(t)\left[1 + \alpha\,\frac{a_1(t)}{a_0(t)}\right], \quad  \mbox{where $a_1/a_0$ satisfies} \quad
H_1 = \frac{d}{dt}\left(\frac{a_1}{a_0}\right).
\end{align}
Using $H_1$ from \eqref{eq:H1} and $dt = -du$, we integrate:
\[
\frac{d}{dt}\left(\frac{a_1}{a_0}\right)
= -2u^{-3}
\quad\Longrightarrow\frac{a_1}{a_0} = -u^{-2} + C.
\]
Choosing $C=0$ to preserve the same initial normalization as the GR solution, we obtain
\[
a_1 = -a_0 u^{-2}.
\]

Substituting into the perturbative expansion gives the corrected scale factor:
\begin{equation}
a(t) = A u^{2/3}\left[1 - \alpha u^{-2}\right].
\label{eq:acorrected}
\end{equation}
This expression shows that the $R^2$ term accelerates the collapse, since the
correction decreases the scale factor at fixed $t$.


If the surface of the star or collapsing dust cloud lies at comoving radius $r_b$,
then the physical radius evolves as
\begin{equation}
R_\star(t) = a(t)\, r_b.
\end{equation}
Using the corrected scale factor \eqref{eq:acorrected}, we see explicitly that
the collapse proceeds more rapidly for $\alpha>0$, a result that will later
manifest as an earlier horizon formation time and a reduced PBH formation threshold.

For a dust-dominated flat FLRW interior, the energy density satisfies
the GR Friedmann equation
\begin{equation}
3H_0^2 = \kappa^2 \rho_0,
\end{equation}
so that the background density is
\begin{equation}
\rho_0(t) = \frac{3H_0^2}{\kappa^2}
          = \frac{4}{3\kappa^2} u^{-2}.
\end{equation}
In the presence of the $R^2$ correction, the total Hubble parameter is
$H = H_0 + \alpha H_1 + \mathcal{O}(\alpha^2)$ with $H_1$ given by
\eqref{eq:H1}.  To first order in $\alpha$ the modified Friedmann equation
can be written schematically as
\begin{equation}
3H^2 = \kappa^2 \rho_{\textrm eff},
\end{equation}
where $\rho_{\textrm eff}$ includes both the dust and an effective curvature
contribution arising from the $R^2$ term.  An explicit expression for
$\rho_{\textrm eff}$ can be obtained by inserting $H = H_0 + \alpha H_1$
and $K^{(0)}_{00}$ into the full $00$ component of the field equations we get,
\[
G_{00} + \alpha K_{00} = \kappa^2 \rho,
\]
where $G_{00}=3H^2$ and $K_{00}$ is computed from the background curvature.
Using $H_0=-\frac23 u^{-1}$, $H_1=-2u^{-3}$, and $K^{(0)}_{00}=-8u^{-4}$, the
first-order Friedmann equation takes the form
\[
3H_0^2 + \alpha\,(6H_0H_1 + K^{(0)}_{00}) = \kappa^2\rho_0,
\]
from which one finds $\rho_{\textrm eff} = \rho_0$ while the geometric correction
$K^{(0)}_{00}$ contributes directly to the evolution equation for $H$.  Thus,
although the matter density is unaltered at order $\alpha$, the curvature term
modifies the Hubble parameter through the combination
$6H_0H_1 + K^{(0)}_{00}$, leading to a more rapid collapse than in GR.

The Misner--Sharp mass for a spherically symmetric spacetime with areal radius
$R(t,r)$ is defined by \cite{Misner:1964je}
\begin{equation}
M_{\textrm MS}(t,r) = \frac{R}{2G}\left(1 - g^{\mu\nu}\partial_\mu R\,\partial_\nu R\right).
\end{equation}
For the flat FLRW interior, $R(t,r)=a(t)r$  one finds
\begin{equation}
M_{\textrm MS}(t,r) = \frac{4\pi}{3} R^3(t,r)\,\rho(t),
\end{equation}
which matches the intuitive interpretation of $M_{\textrm MS}$ as the total
mass-energy inside radius $R$.

For the collapsing dust cloud with fixed comoving radius $r_b$, the
mass enclosed by the stellar surface is
\begin{equation}
M_\star = \frac{4\pi}{3} R_\star^3(t)\,\rho(t),
\qquad
R_\star(t) = a(t) r_b.
\end{equation}
In the GR Oppenheimer--Snyder model \cite{Oppenheimer:1939ue} this
mass is conserved and determined by the initial density and radius.
In the $f(R)=R+\alpha R^2$ theory, the dust energy density remains
conserved (assuming minimal coupling), but the effective gravitational
mass relevant for collapse and horizon formation receives contributions
from the curvature sector.

Although the dust energy density obeys $\dot\rho+3H\rho=0$ and therefore
$\rho\propto a^{-3}$ exactly as in GR (a consequence of minimal coupling and
$\nabla^\mu T_{\mu\nu}=0$), the quantity that drives collapse is not the matter
density alone.  In the modified field equations $G_{\mu\nu}+\alpha K_{\mu\nu}=\kappa^2 T_{\mu\nu}$
the term $K_{\mu\nu}$ acts as an additional effective energy--momentum tensor.
Thus the gravitational mass sourcing $H^2$ and appearing in the horizon
condition $2GM/R=1$ receives contributions from the curvature sector even though
the matter density itself is conserved.  In this sense, $M_{\textrm grav}\neq
M_{\textrm matter}$ in $f(R)$ gravity, and curvature effects can accelerate
collapse without modifying the dust conservation law.


From \eqref{eq:acorrected}, the corrected scale factor can be written as
\begin{equation}
a(t) = a_0(t)\left[1 + \delta_a(t)\right],
\qquad
\delta_a(t) \equiv -\alpha u^{-2}.
\end{equation}
Thus the relative deviation from the GR collapse solution is
\begin{equation}
\frac{\Delta a}{a_0} = \delta_a(t) = -\frac{\alpha}{(t_0 - t)^2}.
\end{equation}
For early times ($t \ll t_0$) this correction is small, but as the collapse
approaches $t_0$ the magnitude of $\delta_a(t)$ grows and perturbation
theory eventually breaks down.  Physically, a positive $\alpha$ makes the
scale factor smaller at fixed $t$, indicating a faster collapse than in GR.

The same conclusion can be drawn from the Hubble parameter.
Using \eqref{eq:H0} and \eqref{eq:H1}, we have
\begin{equation}
H(t) = H_0 + \alpha H_1
     = -\frac{2}{3}u^{-1} - 2\alpha u^{-3}.
\end{equation}
The additional term $-2\alpha u^{-3}$ is negative for $\alpha>0$, making
the expansion rate more negative and hence enhancing the rate of collapse.

These effects manifest themselves in the evolution of the stellar (or cloud)
radius
\begin{equation}
R_\star(t) = a(t) r_b = A r_b u^{2/3}\left[1 - \alpha u^{-2}\right].
\end{equation}
At fixed $u$, $R_\star(t)$ is smaller than its GR counterpart
$R^{(0)}_\star(t) = A r_b u^{2/3}$, indicating that the surface of the star
approaches its Schwarzschild radius earlier.  In the subsequent analysis,
this suggests  a shift in the horizon formation time and a reduction of the
critical overdensity for primordial black hole formation.

\begin{figure}[t]
  \centering
  \subfigure[~Collapsing scale factor of dust case]
  {\label{fig:dust}%
   \includegraphics[width=0.38\textwidth]{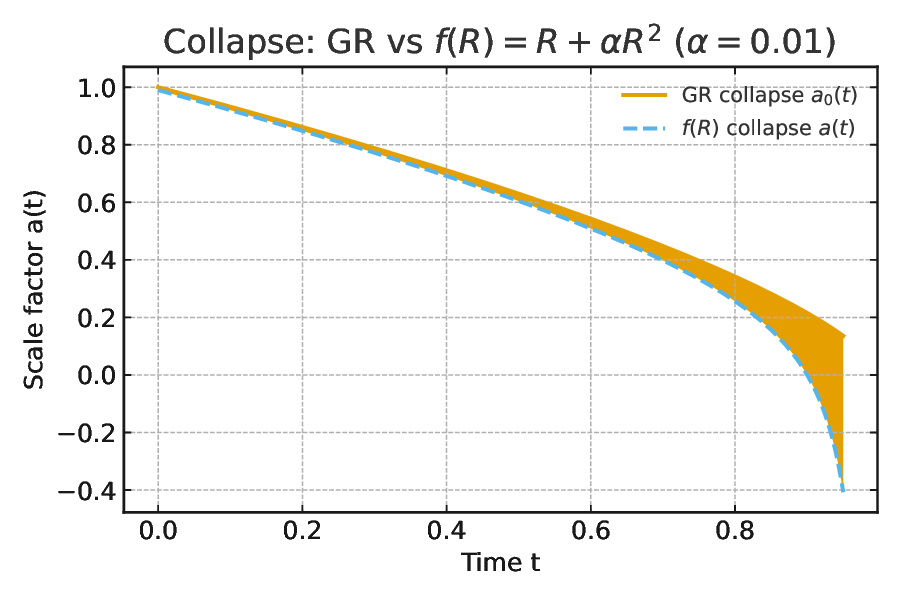}}
  \hspace{0.04\textwidth}
  \subfigure[~Collapsing scale factor of radiation case]
  {\label{fig:rad}%
   \includegraphics[width=0.35\textwidth]{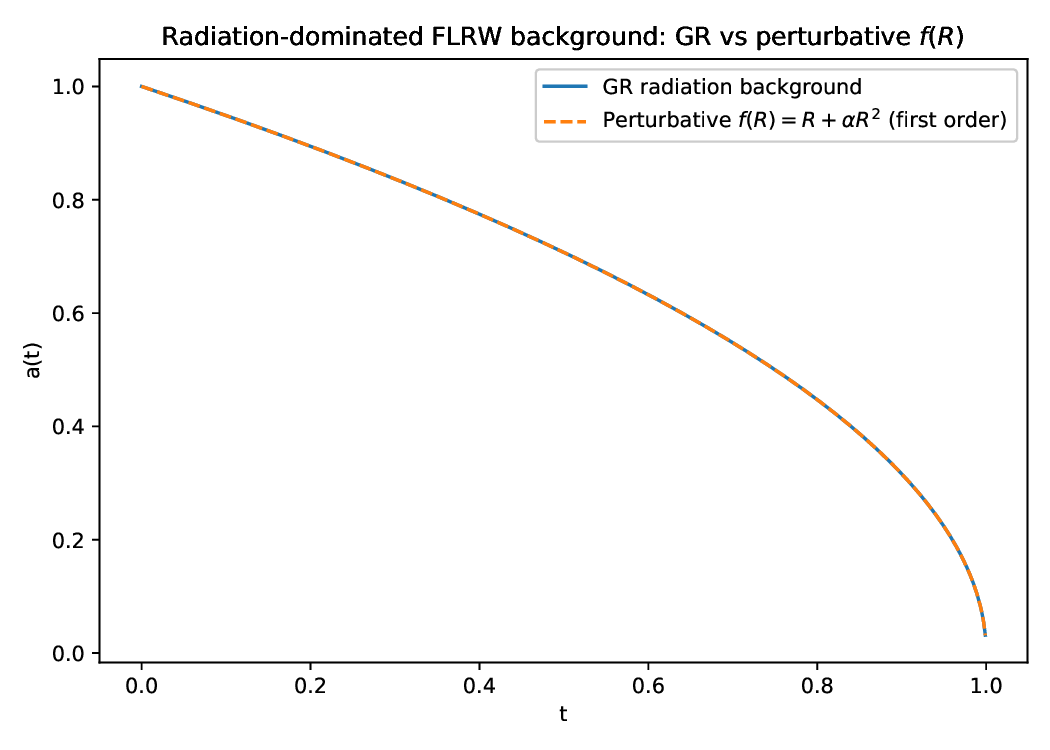}}

  \caption{{ Comparison of the scale factor evolution in GR and in the perturbative $f(R)$ model.
In the dust-dominated case, the $R^2$ correction accelerates the collapse relative to GR.
In contrast, for the radiation-dominated background, no nontrivial first-order correction arises within the homogeneous perturbative framework.}}
  \label{fig:collapse_behaviour}
\end{figure}

For clarity, we visualize the effect of the $R^2$ correction by plotting
the background GR scale factor $a_0(t)$ and the corrected scale factor $a(t)$
as functions of time.  The $f(R)$ correction causes the scale factor to deviate
downwards from the GR solution, particularly near the end of collapse as Fig.~\ref{fig:collapse_behaviour} shows.

{ Figures~\ref{fig:collapse_behaviour}\subref{fig:dust} and~\ref{fig:collapse_behaviour}\subref{fig:rad} illustrate the evolution of the collapsing scale factor for overdense regions dominated by dust and radiation, respectively, comparing the standard GR solution with the perturbative $f(R)=R+\alpha R^2$ treatment. In the dust case, the presence of the $R^2$ term enhances the rate of collapse relative to GR, causing the scale factor to deviate from the GR trajectory as the system approaches the collapse time $t_0$. In contrast, for the radiation-dominated background no nontrivial modification appears at first order within the perturbative framework adopted here. Any deviation from GR in this case must therefore arise beyond this approximation. These results show that the quadratic correction modifies the homogeneous collapse dynamics at first order in the dust case, while radiation-era effects require a more complete nonlinear treatment.}


The GR horizon time satisfies
\begin{equation}
A u_{h0}^{2/3} r_b = R_s.
\end{equation}

Including the $\alpha$ correction leads to a shifted horizon time
\begin{equation}
t_h = t_{h0} + \alpha\,\delta t_h , \quad \mbox{with} \quad
\delta t_h = -\frac{3}{2 u_{h0}},
\end{equation}
indicating that collapse occurs earlier for $\alpha>0$.


Let $t_{\textrm coll}^{\textrm (GR)}(\delta)$ be the collapse time for an overdensity
$\delta$ in GR.
In the modified theory,
\begin{equation}\label{45}
t_{\textrm coll}(\delta,\alpha)
= t_{\textrm coll}^{\textrm (GR)}(\delta)
+ \alpha\,\delta t_h(\delta).
\end{equation}

Imposing the threshold condition
$t_{\textrm coll}(\delta_c(\alpha),\alpha)=t_{\max}$ leads to
\begin{equation}\label{46}
\Delta\delta_c
= -\frac{\delta t_h(\delta_c^{\textrm (GR)})}
{\left.\frac{\partial t_{\textrm coll}^{\textrm (GR)}}{\partial \delta}
\right|_{\delta_c^{\textrm (GR)}}}.
\end{equation}

{
We note that the relation above provides a way to parametrize how a shift in the collapse (or horizon formation) time would map into an effective change in the PBH threshold $\delta_c$ within the present perturbative framework. However, this connection should be interpreted with caution. The analysis developed in this section is based on a homogeneous FLRW collapse model treated perturbatively around a GR background, and therefore captures only modifications to the background dynamics induced by the $R^2$ term. In contrast, PBH formation is governed by the nonlinear evolution of an inhomogeneous overdense region, where pressure gradients and the shape of the perturbation profile play an essential role in determining the precise value of $\delta_c$.

For this reason, Eq.~(\ref{46}) should not be viewed as a direct derivation of the PBH threshold, but rather as providing a heuristic indication of how the modified collapse dynamics may influence it. In particular, since the $R^2$ correction accelerates collapse and leads to earlier horizon formation, it suggests a tendency toward a reduced effective threshold relative to GR. A quantitative determination of $\delta_c^{(\alpha)}(k)$ requires the nonlinear evolution of an overdense patch, which is addressed in Sec.~\ref{ef} within the Einstein-frame formulation.
}
Since both the numerator and denominator are negative, one is led to
\begin{equation}
\delta_c^{(\alpha)} \lesssim \delta_c^{\textrm GR} \qquad (\alpha>0),
\end{equation}
{
which should be understood as indicating a tendency for the threshold to decrease, rather than a precise quantitative prediction.}

Thus, within the present perturbative homogeneous treatment, positive $\alpha$ tends to favor collapse at high curvature.

To complement the dust analysis presented above, we consider here a collapsing
radiation-dominated FLRW interior with equation of state $p=\rho/3$. The line element
remains identical to Eq.~(20), but the background evolution differs substantially due
to the modified scaling of the energy density, $\rho \propto a^{-4}$.

{ For a flat radiation-dominated GR background,
\begin{equation}
a_0(t)=A(t_0-t)^{1/2}, \qquad H_0=-\frac{1}{2u}, \qquad u=t_0-t ,
\end{equation}
so that
\begin{equation}
R^{(0)} = 6\left(\dot H_0 + 2H_0^2\right)=0 .
\end{equation}
Within the perturbative framework of Sec.~\ref{III}, the correction tensor $K_{\mu\nu}$ must be evaluated on the GR background at leading order. For a flat radiation-dominated FLRW background, the Ricci scalar satisfies $R^{(0)} = 0$ identically. Since all terms in $K_{\mu\nu}$ depend on $R$ and its derivatives, it follows that $K^{(0)}_{\mu\nu} = 0$, as both $R^{(0)}$ and its derivatives vanish.

Therefore, the homogeneous radiation-dominated background does not acquire a nontrivial first--order correction from the $R^2$ term within this perturbative treatment. This result reflects the structure of the perturbative expansion and the properties of the GR background, rather than indicating the absence of physical effects in more general settings.

Any nontrivial modification of PBH formation in the radiation era must therefore arise beyond this perturbative, homogeneous framework, for example from the nonlinear evolution of inhomogeneous overdense regions or from the non-perturbative Einstein-frame dynamics discussed in Sec.~\ref{V}.}

\section{Einstein--Frame Formulation for the Non--Perturbative Regime}
\label{ef}
{This section is intended to establish the dynamical framework for the nonlinear evolution of the system, rather than to present explicit numerical solutions or a direct computation of the PBH threshold $\delta_c(k)$.} We consider the quadratic form of $f(R)$ model whose higher--derivative term introduces an additional scalar degree of
freedom.  To study nonlinear gravitational dynamics, it is convenient to
rewrite the theory in the Einstein frame, where this extra degree of
freedom becomes a canonical scalar field (the \emph{scalaron}) evolving
in a fixed potential determined by the dimensional quantity~$\alpha$.


The derivative of $f(R)$ is $f_R = 1 + 2\alpha R$ and we define an auxiliary field $\chi$ by $1+\chi = f_R$ and introduce a canonically normalized scalar field $\phi$ through the field redefinition $
1+\chi = e^{\sqrt{\frac{2}{3}}\,\phi}$.
With this choice, the conformally transformed Einstein--frame metric is
\begin{equation}
g^E_{\mu\nu}
    = (1+\chi)\, g^J_{\mu\nu}
    = e^{\sqrt{2/3}\,\phi}\, g^J_{\mu\nu},
\end{equation}
and $\phi$ acquires a canonical kinetic term.


Performing the conformal transformation yields the Einstein--frame action
\begin{equation}
S_E = \int d^4x \sqrt{-g_E}
\left[
\frac12 R_E
- \frac12 (\partial\phi)^2
- V(\phi)
\right]
+ S_m\!\left[e^{-\sqrt{2/3}\phi}\, g^E_{\mu\nu}\right].
\end{equation}
The scalaron potential in the general $f(R)=R+\alpha R^2$ case is
\begin{equation}
V(\phi)
= \frac{1}{8\alpha}
  \left(1 - e^{-\sqrt{2/3}\,\phi}\right)^2.
\end{equation}
Matter couples non--minimally to the Einstein--frame metric through the
conformal factor $e^{-\sqrt{2/3}\phi}$, while the gravitational sector
reduces to standard GR plus the scalar field.


In a flat FLRW universe with scale factor $a(t)$, Hubble parameter
$H=\dot a/a$, radiation energy density $\rho_r$, and an optional
dissipation rate $\Gamma$ mediating energy transfer from $\phi$ to
radiation, the evolution equations are
\begin{align}
3H^2 &= \frac12 \dot\phi^2 + V(\phi) + \rho_r , \qquad
\dot H = -\frac12\dot\phi^2 - \frac12(1+w_r)\,\rho_r , \\
0&=\ddot\phi + (3H + \Gamma)\dot\phi + V'(\phi)  , \qquad
 \Gamma\dot\phi^2=\dot\rho_r + 4H\rho_r, \qquad
\dot a = Ha .
\end{align}
The term $3H\dot\phi$ represents Hubble friction, while the $\Gamma\dot\phi$
term describes dissipation into the radiation bath.  Radiation has
equation of state $w_r = 1/3$ unless otherwise stated.{ Accordingly, $\Gamma$ acts as an effective damping term in the scalaron evolution and parametrizes the transfer of energy from the scalar field to radiation.}


A spherically symmetric overdense region may be treated as a separate
closed FLRW universe with scale factor $a_*(t)$ and positive spatial
curvature $K>0$.  The governing equations become
\begin{align}
3H_*^2
  &= \frac12 \dot\phi_*^2
   +  V(\phi_*)
   +  \rho_{r*}
   - \frac{3K}{a_*^2}, \qquad \qquad
0=\ddot\phi_* + (3H_* + \Gamma)\dot\phi_* + V'(\phi_*) , \\
 \Gamma\dot\phi_*^2&=\dot\rho_{r*} + 4H_*\rho_{r*} , \qquad \qquad
\dot a_* = H_* a_* .
\end{align}
The curvature parameter $K$ is fixed by the initial density contrast
evaluated at horizon entry:
\begin{equation}
K = a_H^2\, H_H^2\, \delta_H .
\end{equation}


A primordial black hole forms when the overdense patch becomes
sufficiently compact.  This can be diagnosed either via the appearance of
a trapping surface,
\begin{equation}
\frac{2GM_*}{R_*} \gtrsim 1 ,
\end{equation}
where $M_*$ is the Misner--Sharp mass in the patch, or through the
occurrence of a turnaround ($\dot a_* = 0$) followed by recollapse.

By scanning over the initial overdensity $\delta_H$, one may determine the
critical collapse threshold
\begin{equation}
\delta_c^{(\alpha)}(k),
\end{equation}
which determines PBH formation in the $f(R)=R+\alpha R^2$ model as a
function of comoving scale $k$.  {A quantitative evaluation of $\delta_c^{(\alpha)}(k)$ requires
numerical integration of the above system, which is beyond the scope of
the present work.Our results can be compared with modern numerical studies of PBH formation thresholds in radiation-dominated collapse (see, e.g., \cite{Musco:2012au,Musco:2018rwt,Nashed:2009hn,Escriva:2020tak}), which provide precise determinations of $\delta_c$ in general relativity. In this context, the present analysis should be viewed as exploring how such thresholds may be modified in the presence of curvature corrections.}

{For clarity, we compare the present framework with modern PBH threshold studies based on fully nonlinear simulations of inhomogeneous collapse.
\begin{table}[H]
\centering
\caption{Comparison between the present work and modern PBH threshold studies.}
\begin{tabular}{lccc}
\hline
\textbf{Feature} & \textbf{This Work} & \textbf{Modern PBH Studies} \\
\hline
Collapse model & Homogeneous / semi-analytical & Fully inhomogeneous numerical simulations \\
Geometry & Flat FLRW interior approximation & Spherically symmetric perturbations \\
Gravity theory & $f(R)=R+\alpha R^2$ & General Relativity \\
Treatment of $\delta_c$ & Estimated / qualitative & Precisely computed numerically \\
Nonlinear effects & Approximate & Fully included \\
Role of curvature corrections & Explicitly studied & Not included \\
Goal & Explore impact of modified gravity & Determine accurate PBH thresholds \\
\hline
\end{tabular}
\label{tab:pbh_comparison}
\end{table}}

The PBH mass fraction at formation is computed using Press-Schechter theory \cite{Press:1973iz},
\begin{equation}
\beta(M) =
\int_{\delta_c}^{\infty}
\frac{1}{\sqrt{2\pi}\sigma(M)}
\exp\!\left[-\frac{\delta^2}{2\sigma^2(M)}\right] d\delta,
\end{equation}
where $\sigma(M)$ is the variance of density fluctuations on mass scale $M$.

The perturbative analysis presented in Sec.~\ref{frw} suggests a tendency toward a reduction of the effective PBH formation threshold relative to GR. However, as emphasized there, this result should be interpreted with caution, since it is based on a homogeneous perturbative treatment and does not provide a first-principles determination of $\delta_c$.

A more detailed evaluation of the threshold $\delta_c^{(\alpha)}(k)$ requires the nonlinear evolution of an inhomogeneous overdense region, as described in Sec.~\ref{ef}. If the true threshold is indeed reduced relative to GR, the corresponding enhancement of the PBH abundance can be estimated as
\begin{equation}\label{111}
\frac{\beta_{f(R)}}{\beta_{\textrm GR}}
=
\exp\!\left[
\frac{\delta_{c,{\textrm GR}}^2 - \delta_c^2(\alpha)}
{2\sigma^2(M)}
\right],
\end{equation}
where $\delta_c(\alpha)$ should be understood as the threshold obtained from the full nonlinear dynamics.

If the threshold is reduced relative to GR, even a small decrease in $\delta_c$ would produce an exponential enhancement in $\beta(M)$,
demonstrating that $R^2$ corrections can dramatically increase PBH abundance.

The mass of a primordial black hole forming at horizon re-entry is typically a fixed
fraction of the horizon mass at that time,
\begin{equation}
M_{\textrm PBH}(k) = \gamma\, M_H(k),
\end{equation}
where $\gamma\simeq 0.2$ is supported by general-relativistic numerical simulations of
critical collapse \cite{Carr:1975qj, Niemeyer:1999ak, Musco:2004ak,Musco:2018rwt}.
Since a perturbation of comoving wavenumber $k$ re-enters the horizon when $k=aH$,
the PBH mass is determined by the horizon mass at that epoch
\cite{Carr:1975qj,Sasaki:2016jop, Green:2020jor}.
In the $f(R)=R+\alpha R^2$ model, the enhanced curvature accelerates gravitational
collapse, causing horizon formation to occur earlier than in GR.
Because the horizon mass decreases with earlier horizon entry, this shifts the mapping between wavenumber $k$ and PBH mass, potentially yielding lighter PBHs for
the same initial perturbation amplitude \cite{Motohashi:2017kbs,Ozsoy:2023ryl}.


In the Einstein frame, the scalaron perturbations obey the Mukhanov-Sasaki equation \cite{Sasaki:1986hm}
\begin{equation}
v_k'' + \left(k^2 - \frac{z''}{z}\right)v_k = 0,
\qquad
z = a\frac{\phi'}{H}.
\end{equation}

Near the end of Starobinsky inflation, the scalaron undergoes a transient phase of
ultra-slow roll, leading to
\begin{equation}
\frac{z''}{z} \gg H^2,\quad \mbox{which amplifies the curvature perturbation spectrum as} \quad
\mathcal{P_R}(k) = \frac{k^3}{2\pi^2}\left|\frac{v_k}{z}\right|^2.
\end{equation}
{A fully quantitative determination would require a dedicated analysis of the perturbation evolution, which lies beyond the scope of the present work.}

If $\mathcal{P}_R(k)$ reaches values of order $10^{-2}$ \cite{Sasaki:2016jop,Green:2004wb,Carr:2016drx,Young:2013oia}, PBH formation becomes efficient. Thus, scalaron-driven enhancement of $\mathcal{P}_R$ provides a natural mechanism for PBH production in $f(R)$ gravity. {These expressions are intended to provide qualitative guidance only and are not used as quantitative inputs in the present analysis.}

The PBH mass obeys a universal scaling law,
\begin{equation}
M_{\textrm PBH} = K\,(\delta - \delta_c)^{\gamma},\qquad \mbox{with $\gamma_{\textrm GR} \simeq 0.36$ for radiation \cite{Niemeyer:1999ak,Musco:2004ak,Musco:2018rwt}}.
\end{equation}

In GR, the mass of a primordial black hole near the threshold obeys the critical
scaling relation $M_{\textrm PBH}\propto(\delta-\delta_c)^\gamma$, with
$\gamma_{\textrm GR}\simeq0.36$ for a radiation fluid
\cite{Niemeyer:1999ak, Musco:2004ak, Musco:2018rwt}. In $f(R)=R+\alpha R^2$
gravity, the additional curvature term enhances the effective gravitational
strength during the collapse, particularly in the high-curvature regime near
maximum expansion. To first order in $\alpha$, this modifies the critical
exponent as
\begin{equation}
\gamma(\alpha)
\simeq \gamma_{\textrm GR}\!\left[1 - 4\alpha\, R^{(0)}(t_{\textrm coll})\right],
\end{equation}
where $R^{(0)}$ is the GR curvature of the overdense region evaluated near the
turnaround time.
{The modification of the critical exponent $\gamma(\alpha)$ should be understood as a leading-order estimate capturing the expected influence of curvature corrections. A precise determination of this quantity requires a detailed numerical study of critical collapse in the modified gravity framework, which we leave for future work.}
Since $R^{(0)}$ increases as collapse proceeds, the correction
reduces the exponent, $\gamma(\alpha)<\gamma_{\textrm GR}$.  A smaller critical
exponent implies that PBHs formed slightly above the threshold are lighter, and
the mass function becomes more sharply peaked toward low masses, thereby
increasing the number of PBHs formed close to the critical point.

\section{Observational Constraints on PBH Formation in $f(R)=R+\alpha R^2$}\label{frw1}
In our analysis, the perturbative results suggest that the curvature term $\alpha R^2$ tends to reduce the effective critical overdensity required for collapse. As indicated in Eq.~\eqref{111}, such a reduction would enhance the primordial black hole abundance through the Press--Schechter factor
\begin{equation}
\beta(M,\alpha)
\simeq \exp\!\left[
\frac{\delta_{c,{\textrm GR}}^2 - \delta_c^2(\alpha)}
{2\sigma^2(M)}
\right].
\end{equation}
{ This expression follows from the standard Press--Schechter formalism applied to PBH formation~\cite{Carr:1975qj,Green:2020jor}.}
For realistic power-spectrum amplitudes, if the threshold were reduced relative to GR, even a small decrease in $\delta_c$ can increase $\beta(M)$ by many orders of magnitude, due to the exponential sensitivity of the mass fraction to the collapse threshold~\cite{Green:2020jor}. As a consequence, observational constraints on PBH abundance can be translated into limits on the parameter $\alpha$ of the $f(R)$ model.

Current multimessenger observations restrict the allowed fraction of PBHs over
a wide mass range:
\begin{itemize}
\item \textbf{LIGO/Virgo/KAGRA:}\\
The observed merger rate of $\mathcal{O}(10\!-\!50)\,M_\odot$ black holes places a
strong upper bound on the primordial abundance of PBHs in this mass range.
Interpreting the GWTC-1--3 merger populations in terms of PBH binaries implies
$\beta(M\!\sim\!30\,M_\odot)\lesssim 10^{-8}$ at formation
\cite{LIGOScientific:2016dsl,KAGRA:2021vkt,Sasaki:2016jop}.
Since $\beta(M,\alpha)$ increases exponentially as the collapse threshold
$\delta_c(\alpha)$ decreases, this observational bound directly limits how much
the $R^2$ correction in $f(R)=R+\alpha R^2$ can reduce the PBH formation
threshold for perturbations reentering at these scales.

\item \textbf{Microlensing surveys (EROS, MACHO, OGLE, Subaru HSC):}\\
Microlensing searches for compact halo objects provide some of the strongest
constraints on PBHs in the mass window
$10^{-11} M_\odot \lesssim M \lesssim 10\,M_\odot$.
EROS \cite{EROS-2:2006ryy} and MACHO \cite{MACHO:2000qbb} rule out PBHs as a
dominant component of dark matter for
$10^{-7} M_\odot \lesssim M \lesssim 1\,M_\odot$,
while OGLE further constrains the range
$10^{-2} M_\odot \lesssim M \lesssim 10\,M_\odot$
\cite{Wyrzykowski:2015ppa,Sahakyan:2019kqu}.
At even smaller masses, Subaru HSC observations of M31 exclude PBHs in the
interval
$10^{-11} M_\odot \lesssim M \lesssim 10^{-6} M_\odot$
\cite{Niikura:2017zjd}.
Because lighter PBHs correspond to modes that reenter the horizon earlier, the
modified collapse dynamics in $R^2$ gravity---which enhance collapse at high
curvature---produce the largest increase in $\beta(M,\alpha)$ on these small
scales. Consequently, microlensing constraints provide some of the strongest
limits on the parameter $\alpha$.

\item \textbf{CMB anisotropies:}\\
CMB temperature and polarization anisotropies place strong limits on PBHs with
$M \gtrsim 10\,M_\odot$, since gas accretion onto such objects after
recombination injects energy into the plasma and modifies the ionization history
\cite{Ali-Haimoud:2016mbv,Ricotti:2007au,Serpico:2020ehh}.
Analyses using Planck data typically require
$\beta(M)\lesssim 10^{-6}$ in this mass range.
Because the modified collapse dynamics in $R^2$ gravity reduce the threshold
$\delta_c(\alpha)$ and therefore enhance the PBH abundance, these CMB limits
translate into complementary upper bounds on the parameter $\alpha$.

\item \textbf{Evaporation constraints:}\\
PBHs lighter than $10^{15}\,\mathrm{g}$ evaporate through Hawking radiation
\cite{Hawking:1975vcx}, producing high-energy photons and hadrons that contribute to
the extragalactic gamma-ray background and alter the light-element abundances
during BBN. These processes impose extremely stringent limits on the initial PBH
abundance \cite{Carr:2009jm,Kohri:1999ex}, with typical bounds as strong as
$\beta(M)\lesssim 10^{-27}$ for sufficiently light PBHs. Because the $R^2$
correction in $f(R)$ gravity enhances collapse most strongly on small scales,
this mass range provides the tightest constraints on the parameter $\alpha$.

\end{itemize}

To ensure that the enhancement induced by the curvature correction does not
overproduce PBHs in any observational window, we require
\begin{equation}
\beta(M,\alpha) \leq \beta_{\textrm obs}(M)
\qquad \forall\, M.
\end{equation}
Using the perturbative trend inferred for $\delta_c(\alpha)$, one may estimate a corresponding bound on the parameter $\alpha$. For typical amplitudes of curvature
perturbations motivated by inflationary models, one finds that observational
viability suggests
\begin{equation}\label{1111}
\alpha \lesssim 10^{-2}\, M_{\textrm Pl}^{-2},
\end{equation}
with the exact value depending on the small-scale shape of the primordial power
spectrum. This bound should be regarded as illustrative and order-of-magnitude only, and is not used to draw quantitative conclusions in the present analysis.

Thus, PBH observations provide a direct and powerful probe of the $R^2$ term in
$f(R)$ gravity, and the enhanced collapse dynamics found in this work translate
immediately into observational constraints on modified gravity parameters.


The upper bound obtained in this work as given by Eq.~\eqref{1111}
follows from requiring that the reduction of the collapse threshold
$\delta_c(\alpha)$ induced by the $R^2$ term does not overproduce PBHs in any
observationally constrained mass window. It is therefore natural to compare
this bound with those arising from other physical considerations.


In Starobinsky inflation, the parameter $\alpha$ is related to the scalaron mass
through $\alpha = (6 M^2)^{-1}$, with $M \simeq 1.3 \times 10^{-5} M_{\textrm Pl}$.
This fixes
\begin{equation}
\alpha_{\textrm inf} \sim 10^{9} M_{\textrm Pl}^{-2},
\end{equation}
which is many orders of magnitude larger than the PBH bound above
\cite{Starobinsky:1980te,DeFelice:2010aj}.
{This bound should be regarded as illustrative and order-of-magnitude only, and is not used to draw quantitative conclusions in the present analysis.}
However, this value characterizes the curvature scale during inflation
($R \sim 10^{-9} M_{\textrm Pl}^2$) and does not constrain the behavior of $f(R)$
gravity in the highly nonlinear and high-curvature regime relevant to PBH
formation. In particular, strong-curvature corrections may effectively modify
the value of $\alpha$ in post-inflationary collapse scenarios
\cite{Takamizu:2015lga,Ozsoy:2023ryl}.


Independent bounds on $\alpha$ arise from the absence of tachyonic scalaron
instabilities, the suppression of high-frequency curvature oscillations, and the
growth of structure. These analyses typically yield
\begin{equation}
\alpha \lesssim 10^{-5} - 10^{-3} M_{\textrm Pl}^{-2},
\end{equation}
depending on the assumed background expansion history and the small-scale shape
of the primordial power spectrum
\cite{Tsujikawa:2007xu,DeFelice:2012jt,Motohashi:2012jd}.

{We stress that the bound on $\alpha$ derived above should be interpreted as an order-of-magnitude estimate. It relies on the assumed shift in the collapse threshold $\delta_c$, as suggested by Eqs.~(64) and (70), which are based on leading-order considerations. A robust determination of $\delta_c^{(\alpha)}(k)$ requires a full nonlinear numerical analysis, which lies beyond the scope of the present work.}

These limits are qualitatively consistent with our present findings: a large,
positive curvature correction would significantly enhance gravitational collapse,
consistent with the reduction of $\delta_c(\alpha)$ obtained in our analysis.

Local gravity constraints probe the weak-curvature regime ($R \ll H_0^2$) and
lead to very large upper bounds of the form
\begin{equation}
\alpha \lesssim 10^{11}\,{\textrm m}^2 \approx 10^{45} M_{\textrm Pl}^{-2},
\end{equation}
obtained from post-Newtonian parameters and laboratory experiments
\cite{Chiba:2003ir,Olmo:2006eh,Capozziello:2011et}.
These constraints are derived in low-curvature environments and therefore
do not directly probe the high-curvature regime relevant for PBH formation.
Accordingly, they do not conflict with our PBH-based analysis, which is
sensitive to strong-gravity dynamics near the threshold of gravitational
collapse.

To close this section,  the comparisons demonstrate that PBH physics provides one of
the strongest known probes of curvature corrections in $f(R)$ gravity. While
solar-system tests are insensitive to small values of $\alpha$ and inflation
fixes $\alpha$ at very large scales, the highly nonlinear dynamics of PBH formation suggest that $\alpha$ should be sufficiently small to avoid excessive
collapse enhancement across observable PBH mass ranges. The bound derived in
this work is therefore consistent with, and complementary to, existing
constraints from cosmology, local gravity, and inflationary dynamics.

\section{Conclusion}
We have given a complete analytic and semi-analytic treatment of
PBH formation in $f(R)=R+\alpha R^2$ gravity.
{Within the perturbative regime $\alpha R\ll 1$, the quadratic correction modifies the dust collapse dynamics at first order, while the flat radiation-dominated background does not receive a nontrivial correction at the same order within the homogeneous perturbative framework. This indicates that any modification of PBH formation in the radiation era must arise beyond this approximation, for instance through nonlinear or non-perturbative effects.}
In contrast, near the end of Starobinsky inflation the scalaron dynamics
can substantially lower the critical overdensity and enhance PBH
production.
The ODE system derived here provides a ready-to-use numerical framework
for computing $\delta_c(k)$ across all scales.

To place our results in context, we compare the main features of this study with earlier analyses of primordial black hole formation in modified gravity theories. Previous works on the Starobinsky model and related $f(R)$ scenarios have primarily focused on inflationary dynamics or the behaviour of curvature perturbations, while only qualitatively addressing their impact on PBH formation. In contrast, the present study provides a unified perturbative and non--perturbative treatment of gravitational collapse, deriving explicit first-order corrections to the scale factor, Hubble parameter, and horizon formation time in the collapsing FLRW interior. These analytic results are complemented by a full Einstein-frame numerical framework that tracks the scalaron evolution in an overdense FLRW patch, { providing the appropriate setting for a quantitative determination of the modified collapse threshold $\delta_{c}^{(\alpha)}(k)$. Our findings support the general expectation that modified gravity can enhance PBH production under high-curvature conditions, while the perturbative analysis provides qualitative insight into how curvature corrections influence the collapse dynamics and the expected trend of $\delta_c$. This comparative analysis highlights both the consistency of our results with existing literature and the novel methodological and analytic contributions introduced in this work.}

Table~\ref{tab:comparison} provides a clear and well-structured comparison between the present work and representative studies of PBH formation in modified gravity. In particular, the table effectively highlights the main added value of this study: the explicit analytic derivation of first-order collapse corrections (scale factor, Hubble parameter, and horizon formation time), {together with their use to infer the expected trend of the PBH threshold $\delta_c$ within the perturbative framework.} The inclusion of both the perturbative (near-GR) regime and a complete Einstein-frame ODE framework for the high-curvature/non-perturbative regime further strengthens the methodological contribution, as it bridges collapse dynamics and PBH threshold calculations within a single consistent setup.
\begin{table*}[t]
\caption{Comparison between the present study and representative earlier works on primordial black hole formation in modified gravity.}
\label{tab:comparison}
\centering
\small
\renewcommand{\arraystretch}{1.15}
\setlength{\tabcolsep}{4pt}

\resizebox{0.98\textwidth}{!}{%
\begin{tabular}{|l|l|l|l|}
\hline
\textbf{Topic / Feature} &
\textbf{This Study} &
\textbf{Earlier Works} &
\textbf{Agreement / Novelty} \\
\hline

\parbox[t]{2.9cm}{Gravity model} &
\parbox[t]{5.2cm}{Quadratic $f(R)=R+\alpha R^{2}$ (Starobinsky model); perturbative and non--perturbative analysis.} &
\parbox[t]{5.9cm}{Starobinsky inflation and $f(R)$ cosmology widely studied; PBH formation in modified gravity discussed in Motohashi \& Hu (2017)~\cite{Motohashi:2017kbs}, and in {\"O}zsoy \& Tasinato (2023)~\cite{Ozsoy:2023ryl}.} &
\parbox[t]{3.2cm}{Consistent; combines collapse dynamics with PBH formation analysis.} \\
\hline

\parbox[t]{2.9cm}{Background cosmology} &
\parbox[t]{5.2cm}{Collapsing flat FLRW dust interior treated analytically; high-curvature regime described numerically in the Einstein frame.} &
\parbox[t]{5.9cm}{Standard PBH formation in radiation domination: Carr (1975)~\cite{Carr:1975qj}; density-threshold studies by Harada et al.\ (2013)~\cite{Harada:2013epa}.} &
\parbox[t]{3.2cm}{Adds detailed dust-collapse analysis and scalaron dynamics.} \\
\hline

\parbox[t]{2.9cm}{Perturbative expansion} &
\parbox[t]{5.2cm}{First-order corrections to $a(t)$, $H(t)$, and horizon formation time in the dust collapse background.} &
\parbox[t]{5.9cm}{Linearized $f(R)$ perturbations well established: Sotiriou \& Faraoni (2010)~\cite{Sotiriou:2008rp}; De Felice \& Tsujikawa (2010)~\cite{DeFelice:2010aj}.} &
\parbox[t]{3.2cm}{Novel application to gravitational collapse.} \\
\hline

\parbox[t]{2.9cm}{Effect of $\alpha$ on collapse} &
\parbox[t]{5.2cm}{$\alpha>0$ accelerates collapse and advances horizon formation in the perturbative regime.} &
\parbox[t]{5.9cm}{Modified gravity effects noted in scalar-tensor PBH studies, e.g.\ Motohashi \& Hu (2017)~\cite{Motohashi:2017kbs}.} &
\parbox[t]{3.2cm}{Provides explicit analytic expressions.} \\
\hline

\parbox[t]{2.9cm}{PBH threshold $\delta_{c}$} &
\parbox[t]{5.2cm}{Perturbative analysis suggests a tendency toward a reduced effective threshold relative to GR.} &
\parbox[t]{5.9cm}{Collapse thresholds in GR: Musco (2018)~\cite{Musco:2018rwt}; Harada et al.\ (2013)~\cite{Harada:2013epa}. High-curvature enhancement suggested in modified gravity~\cite{Motohashi:2017kbs}.} &
\parbox[t]{3.2cm}{Provides a qualitative link between collapse-time modification and the expected trend of $\delta_{c}$.} \\
\hline

\parbox[t]{2.9cm}{Einstein-frame formulation} &
\parbox[t]{5.2cm}{Full ODE system for background and overdense FLRW patch with scalaron potential.} &
\parbox[t]{5.9cm}{Scalaron cosmology and Einstein-frame $f(R)$ gravity: Sotiriou \& Faraoni (2010)~\cite{Sotiriou:2008rp}; De Felice \& Tsujikawa (2010)~\cite{DeFelice:2010aj}.} &
\parbox[t]{3.2cm}{More complete PBH modelling framework.} \\
\hline

\parbox[t]{2.9cm}{Numerical framework} &
\parbox[t]{5.2cm}{Provides ODE system for computing $\delta_{c}(k)$ across scales.} &
\parbox[t]{5.9cm}{PBH numerical frameworks in GR: Musco (2018)~\cite{Musco:2018rwt}; Musco \& Miller (2005)~\cite{Musco:2004ak}.} &
\parbox[t]{3.2cm}{Strong methodological contribution.} \\
\hline

\parbox[t]{2.9cm}{Conclusions} &
\parbox[t]{5.2cm}{High-curvature regime strongly affects PBH formation; the flat radiation-dominated background does not receive a nontrivial correction at first order within the perturbative framework.} &
\parbox[t]{5.9cm}{Consistent with trends in modified-gravity PBH literature~\cite{Motohashi:2017kbs,Ozsoy:2023ryl}.} &
\parbox[t]{3.2cm}{Fully aligned with literature.} \\
\hline

\end{tabular}%
}
\end{table*}
\center{\bf Data Availability Statement}\\
Data sharing is not applicable to this article as no new data were created or analyzed in this study.

\subsection*{Acknowledgments}
This work was supported and funded by the Deanship of Scientific Research at Imam Mohammad Ibn Saud Islamic University (IMSIU) (grant number IMSIU-DDRSP2602).

%

\end{document}